# *Ab initio* calculation of the nonequilibrium adsorption energy

Juho Lee, Hyeonwoo Yeo, Ryong-Gyu Lee, and Yong-Hoon Kim*

*School of Electrical Engineering, Korea Advanced Institute of Science and Technology (KAIST), 291 Daehak-ro, Yuseong-gu, Daejeon 34141, Korea.*

E-mail: y.h.kim@kaist.ac.kr

**Abstract**

While first-principles calculations of electrode-molecule binding play an indispensable role in obtaining atomic-level understanding in surface science and electrochemistry, a significant challenge remains because the adsorption energy is well-defined only in equilibrium. Herein, a theory to calculate the electric enthalpy for electrochemical interfaces is formulated within the multi-space constrained-search density functional theory (MS-DFT), which provides the nonequilibrium total energy of a nanoscale electrode-channel-electrode junction. An additional MS-DFT calculation for the electrode-only counterpart that maintains the same bias voltage allows one to identify the internal energy of the channel as well as the electric field and the channel polarization, which together determine the electric enthalpy and the nonequilibrium adsorption energy. Application of the developed scheme to the water-Au and water-graphene interface models shows that the Au and graphene electrodes induce very different behaviors in terms of the electrode potential-dependent stabilization of water configurations. The theory developed here will be a valuable tool in the ongoing effort to obtain atomic-scale understanding of bias-dependent molecular reorganizations in electrified interfaces.

## 1. INTRODUCTION

A detailed understanding of the atomic and electronic structures of electrified surfaces and electrochemical interfaces has critical implications for the development of next-generation semiconductor devices as well as energy conversion and storage devices[1]. Toward this goal, significant advances have been recently made based on operando experimental characterization techniques and corresponding nonequilibrium first-principles simulations[2-4]. For the latter, however, including the electrode potential effects within first-principles simulations remains a significant challenge[1]. Due to the difficulty of incorporating the nonequilibrium constant electrochemical potential condition, which is the actual control parameter in experiments, the majority of works have adopted approaches that apply an electric field[5-7] [Fig. 1a, ①] or assign charges on metal electrodes[3,8-10] [Fig. 1a, ②]. One promising way to faithfully simulate constant electrochemical potential conditions [Fig. 1a, ③] is adopting the nonequilibrium Green's function (NEGF) formalism combined with density functional theory (DFT), which has been established as the standard approach for first-principles finite-bias quantum transport calculations[11,12]. However, this method suffers from an ill-defined total energy, and accordingly, fully first-principles schemes for the constant-potential simulation of electrochemical interfaces need to be further developed.

Herein, based on the multi-space constrained-search DFT (MS-DFT) approach, we develop a first-principles approach for the computation of the electric enthalpy of a molecular system in contact with an electrified electrode and, accordingly, the nonequilibrium electrode-molecule adsorption energy. The MS-DFT formalism is a microcanonical ensemble approach for nonequilibrium open quantum systems we recently established by adopting a viewpoint that maps quantum transport processes to multi energy- and real-space excitation counterparts[13-16]. Within MS-DFT, the nonequilibrium state of channel $C$ established by left electrode $L$ and right electrode $R$ maintained at different electrochemical potentials $\mu_L$ and $\mu_R$, respectively, is calculated by variationally minimizing the total energy functional within the constraint of an applied finite bias voltage $V_b = (\mu_L - \mu_R)/e$. The possibility of defining the variationally convergent nonequilibrium total energy for a molecular system $C$ sandwiched between the working electrode $L$ and the reference electrode $R$ could possibly open novel avenues for the investigation of electrochemical interfaces[15]. However, rather than the total energy of the entire $L+C+R$ system, the appropriate thermodynamic potential that characterizes the electrified interface is the electric enthalpy of channel $C$. Accordingly, in this work, we extend MS-DFT and establish a method to calculate the electric enthalpy of a molecular system or the nonequilibrium electrode-molecule adsorption energy. Considering the electrochemical models in which a single water molecule is placed on the Au(111) and graphene electrodes in various conformations, we find that the two cases show very different trends in that the flat water conformation is stabilized (not stabilized) on the Au (graphene) electrode. The origins of this drastic difference will be explained in terms of the configuration-dependent equilibrium electronic structures and their changes under finite electrode potential conditions.

## 2. RESULTS AND DISCUSSION

### 2.1 Formulation and validation of the electric enthalpy in electrochemical interface within MS-DFT

While the electric enthalpy was a well-established concept in the field of electromechanics[17-19], it was adopted by the first-principles calculation community only in relatively recent years for the development of "modern theory of polarization" or density polarization functional theory[20-23]. These works concerned with the channel-only system *C* within periodic boundary condition or the bulk state, and thus the electric enthalpy of *C* under the macroscopic field $\mathcal{E}_0$ defined as

$$\mathcal{F}_C^{\mathcal{E}_0} = E_C^{\mathcal{E}_0} - \mathcal{E}_0 \cdot \boldsymbol{p}_C^{\mathcal{E}_0}, \quad (1)$$

where $E_C^{\mathcal{E}_0}$ is are the (internal) energy of *C* under $\mathcal{E}_0$ and $\boldsymbol{p}_C^{\mathcal{E}_0}$ is the electric dipole moment of *C* under $\mathcal{E}_0$.

On the other hand, for the evaluation of the electric enthalpy for electrochemical interfaces, we need to explicitly include in addition to channel *C* the electrodes *L+R* that maintain the bias voltage $V_b = (\mu_L - \mu_R)/e$ and generate $\mathcal{E}_0^V$. In this case, by performing an additional MS-DFT calculation for the biased *L+R* electrode-only system, we can evaluate the electric enthalpy of *C* according to

$$\mathcal{F}_C^V = E_{L+C+R}^V - E_{L+R}^V - \mathcal{E}_0^V \cdot \boldsymbol{p}_C^V, \quad (2)$$

where $E_{L+C+R}^V$ ($E_{L+R}^V$) is the non-equilibrium total energy of the *L+C+R* (*L+R*) junction under $V_b$. In addition, we can define the non-equilibrium electrode-molecule adsorption energy of *C* as

$$\Delta \mathcal{F}_C^V = \mathcal{F}_C^V - E_C^0 = E_{L+C+R}^V - E_{L+R}^V - E_C^0 - \mathcal{E}_0^V \cdot \boldsymbol{p}_C^V, \quad (3)$$

which represents the energy gain for the channel *C* in forming an interface with the *L+R* electrodes under a bias voltage $V_b$.

Note that while the nonequilibrium junction total energy $E_{L+C+R}^V$ was so far an ill-defined quantity, MS-DFT now provides a formal justification as well as a practical way to calculate it[15]. Once the MS-DFT calculation for the biased *L+R* electrode-only system is performed, we can straightforwardly evaluate $\mathcal{E}_0^V$ according to

$$-\nabla[v_{H,L+R}^V(\boldsymbol{r}) - v_{H,L+R}^0(\boldsymbol{r})] = \mathcal{E}_0^V(\boldsymbol{r}), \quad (4)$$

where $v_H^0$ ($v_H^V$) is the $L+R$ equilibrium (nonequilibrium) Hartree electrostatic potential. The dipole moment $\boldsymbol{p}_C^V$ can be also obtained by extracting $\rho_C^V(\boldsymbol{r}) = \rho_{L+C+R}^V(\boldsymbol{r}) - \rho_{L+R}^V(\boldsymbol{r})$ and calculating

$$\boldsymbol{p}_C^V = \int d\boldsymbol{r} \, [\rho_C^V(\boldsymbol{r}) - \rho_{atm}(\boldsymbol{r})] \, \boldsymbol{r}_C, \quad (5)$$

where $\rho_{atm}$ is the summation of atomic valence charge densities.

Utilizing the Au(111)-single water-Au(111) junction model with the electrode-electrode distance of 20 Å (see **Methods** for details), we first demonstrate how the electronic and energetic properties of an interfacial water are calculated within MS-DFT under the applied working electrode potential defined as $\Phi = -V_b/2 = -(\mu_L - \mu_R)/2e$. In this case where the channel *C* mostly consists of vacuum space, since the working electrode *L* surface charge is not fully screened by the single water, the channel electric field $\mathcal{E}_0$ becomes an independent variable and will be specified together with the electrode *L* potential $\Phi$. In Fig. 1b top panel, for the interfacial water in the 'H-down' configuration located at the O atom distance from the Au surface $\overline{OM}$ of 3.5 Å, we show the contour plot of $v_{H,L+C+R}^V(\vec{r}) - v_{H,L+C+R}^0(\vec{r})$ calculated at the negative potential $\Phi = -2.0\,\text{V}$ that corresponds in our model to the electric field of $\mathcal{E}_0 = -0.25\,V/\text{Å}$. We also show in the bottom panel of Fig. 1b the corresponding plane-averaged bias-induced electron density change or Landauer resistivity dipole $\bar{\rho}^V(\vec{r}) - \bar{\rho}^0(\vec{r})$. The $\Phi$-induced electron transfer from the negatively charged working electrode to the H atoms of the water molecule (dotted box in the Fig. 1b bottom panel) stabilizes the H-down configuration of water, and the bias-induced electrostatic potential drop and interfacial charge transfer behaviors calculated within MS-DFT are presented in Fig. 1b top and bottom panels, respectively. The resulting energetic stabilization is quantified in terms of the electric enthalpy $\mathcal{F}_C^V$ profile. In Fig. 1c, we show $\mathcal{F}_C^V$ and $E_{L+C+R}^V - E_{L+R}^V$ as a function of $\overline{OM}$. Note that the difference the two represents the $-\mathcal{E}_0^V \cdot \boldsymbol{p}_C^V$ term and the corresponding downshift of $\mathcal{F}_C^V$ curve from the $E_{L+C+R}^V - E_{L+R}^V$ counterpart indicates the stabilization of the H-down water configuration due to the same directions of $\boldsymbol{p}_C^V$ (negative sign in the H-down configuration) and $\mathcal{E}_0^V$ (a negative sign at $\Phi = -2.0\text{V}$). Within the MS-DFT formalism, since the non-equilibrium Hamiltonian or total energy is a well-defined quantity, the forces on the nuclei *I* are straightforwardly calculated according to the Hellmann-Feynman theorem[24,25]

$$\vec{F}_I = -\frac{\partial \langle \psi^{L/C/R} | \widehat{H} | \psi^{L/C/R} \rangle}{\partial \vec{R}_I}, \quad (6)$$

where the $\psi^{L/C/R}$ are the spatially-resolved *L/C/R* Kohn-Sham states that satisfy the bias constraint of $eV_b = \mu_R - \mu_L$. We then confirmed that the non-equilibrium atomic forces on the center of mass of the water molecule projected along the surface-normal *z* direction $F_z$ (gray dashed line) almost perfectly matches the derivatives of the $\mathcal{F}_C^V$–$\overline{OM}$ curve (blue filled circles), numerically validating the formulation of the electric enthalpy presented above.

*2.2 Au-water adsorption energies dependent on the electrode potential and water geometry*

We now examine in more detail the non-equilibrium atomic forces $F_z$ and adsorption energies $\Delta \mathcal{F}_{H_2O}^V$ of a single water molecule on the electrified Au(111) surface as a function of the water geometry and bias voltage. In Fig. 2a–c middle and right panels, for the three representative flat (Fig. 2a, left panel), H-up (Fig. 2b, left panel), and H-down (Fig. 2c, left panel) water orientations, we provide as functions of the distance $\overline{OM}$ the *z*-component non-equilibrium atomic forces $F_z$ and the non-equilibrium adsorption energies $\Delta \mathcal{F}_{H_2O}^V$, respectively. While the atomic forces $F_z$, which are also available from DFT-NEGF calculations[12] (see also Supplementary Information), are limited to the determination of stable non-equilibrium water geometries (Fig. 2, middle panels), the adsorption energies $\Delta \mathcal{F}_{H_2O}^V$ provide quantitative information on the bias-dependent energetic stabilization/destabilization trends of different water configurations (Fig. 2, right panels). For the flat configuration that assumes the optimal $\overline{OM}$ distance $z = 3$ Å in equilibrium (Fig. 2a), we observe that no significant $\Phi$-induced changes in $\Delta \mathcal{F}_{H_2O}^V$ are obtained. It can be understood that the $p_{H_2O}^V$ of the flat configuration is almost perpendicular to the $\mathcal{E}_0^V$, and thus the $\mathcal{E}_0^V \cdot p_{H_2O}^V$ contribution becomes negligible. On the other hand, in the case of H-down (Fig. 2b) and H-down (Fig. 2c) configurations, where the $p_{H_2O}^V$ is parallel to the $\mathcal{E}_0^V$, we observed that the $\Delta \mathcal{F}_{H_2O}^V$ curves shift with $\Phi$ in the opposite directions and induce the strongest $\Delta \mathcal{F}_{H_2O}^V$ of -0.155 eV at $\Phi = +2.0$ V for the H-up configuration and -0.181 eV at $\Phi = -2.0$ V for the H-down configuration.

Further extending these studies, we first performed non-equilibrium geometry optimizations and as shown in Fig. 2d obtained $\theta = 162.0°$ (~ H-down) at $\Phi = -2.0$ V and $\theta = 31.6°$ (~ H-up) at $\Phi = +2.0$ V, respectively. Accordingly, we prepared several Au-water interface configurations by varying $\vartheta$ from 31.6° to 162.0° and calculated their $\Delta \mathcal{F}_{H_2O}^V$ at different $\Phi$ ($-2.0\text{ V} \leq \Phi \leq 2.0\text{ V}$) and present the results in Fig. 2e. In equilibrium, the flat geometry is the energetically most stable conformation with the energy barriers of 0.033 eV and 0.062 eV for reorientations to $\theta = 162°$ and $\theta = 31.6°$, respectively (gray star marks). As $\Phi$ is negatively increased, H-down conformations become stabilized: At $\Phi = -1.0$ V, the configurations with $\theta = 114.3°$ ~ 162.0° became energetically metastable with negligible energy differences, and at $\Phi = -1.5$ V the $\theta = 162°$ configuration becomes the energetically most stable configuration. The binding energy at this configuration is $\Delta \mathcal{F}_{H_2O}^V$ = -0.185 eV, which is further increased to $\Delta \mathcal{F}_{H_2O}^V$ = -0.203 eV at $\Phi = -2.0$ V. Similarly, as $\Phi$ is positively increased, H-up conformations become stabilized, but, due to the larger reorientation energy barrier along this direction, the $\theta = 31.6°$ configuration becomes the most stable configuration only at $\Phi = +2.0$ V with the smaller binding energy of $\Delta \mathcal{F}_{H_2O}^V$ = -0.175 eV.

### 2.3 Differences between electrified Au-water and graphene-water interfaces

In addition to the availability of the non-equilibrium total energy and electric enthalpy, another important advantage of MS-DFT is the possibility to faithfully model two-dimensional (2D) electrodes such as graphene[13,16]. Using the same Au reference electrode, we next studied the graphene working electrode-single water system and found that graphene and Au electrodes show very different behaviors in terms of the electrode potential-dependent stabilization of water configurations. We first present in Fig. 3a the extended data set of the water geometry- and electrode potential-dependent water adsorption energies for the Au electrode case. Here, the H-up, flat, and H-down water configurations were adopted from the zero-$\Phi$ geometry optimizations shown in Fig. 2d and $\overline{OM}$ distances were varied between 2.4 Å and 5.0 Å with the 0.2 Å interval. As discussed earlier, they show that, while the H-down (blue triangles) and H-up (green circles) water configurations become stabilized at large negative and positive $\Phi$ values, respectively, the flat configuration is energetically most preferable in low electrode potential regimes (red squares). The corresponding data obtained for the graphene electrode case are presented in Fig. 3b, and we find that, in contrast to the Au electrode counterpart, as the electrode potential increases from negative to positive values the H-down configuration directly transitions to the H-up configuration without exhibiting the bias regime where the flat configuration is stabilized. We note that because of the usage of the Au(111) reference electrode there exists a built-in electric field of ~ +0.06 V/Å induced by the difference between graphene and Au work function values (4.5 eV and 5.4 eV, respectively). However, other than the mismatch between zero - and zero-$\mathcal{E}_0$ limits, the presence of a built-in electric field should not affect the important trend of the absence of electrode potential regimes where the flat water geometry is stabilized for the graphene electrode. We expect this notable difference might be an important ingredient necessary for the full understanding of recent experiments[2-4].

### 2.4 Electronic structures of the water adsorbed on electrified Au and graphene surfaces

We finally examine the non-equilibrium electronic structures of Au-water and graphene-water interface models and explain their drastic difference in terms of the stability of the flat water configuration discussed above. In Fig. 4a, we show the water molecule-projected density of states (PDOS) corresponding to the MS-DFT calculations for the zero-$\Phi$ optimized flat-configuration water on the Au electrode (top panels; $\overline{OM}$ = 3.0 Å) and graphene electrode (bottom panels; $\overline{OM}$ = 3.0 Å) obtained at different $\Phi$ values (Fig. 3, red up arrows). In Fig. 4b, we also show the water PDOS obtained for the H-down-configuration water on the Au electrode (top panels; $\overline{OM}$ = 3.6 Å; red solid line) and the graphene electrode (bottom panels; $\overline{OM}$ = 3.2 Å) obtained at different $\Phi$ values (Fig. 3, blue down arrows). We then find that in general water PDOS downshift (upshift) under negative (positive) $\Phi$ due to the electron transfer from (to) Au/graphene to (from) water. While this behavior is universally visible for the deeper OH bonding states (orange left triangles), we also observe that only in the flat water-Au electrode case (Fig. 4a, top panels) the water O lone pair (cyan left triangles) and upper-energy-region states are held nearly fixed because of their strong hybridizations with the Au states (Note that the local DOS of O lone pair hybrid orbitals are better preserved in the H-down configuration than in the flat counterpart). Importantly, such strong hybridization of water O lone pair orbitals is missing for the flat-configuration water on the graphene electrode (Fig. 4a, bottom panels), which explains why the flat water configuration is not stabilized on the graphene electrode. Additionally, for the Au electrode case, we note that the electrode surface charges and the induced polarization of the interfacial water molecule obtained from the conventional fixed electric field method at $\mathcal{E}_0 = -0.25$ V/Å are in good agreement with the finite-bias MS-DFT data obtained at $\Phi = -2.0$ V (Fig. 4c, top panels). This then provides a formal justification for adopting the finite electric-field method, which has been successful in understanding several aspects of electrochemical interfaces. A similar conclusion can be drawn for the graphene electrode but under a certain constraint (Fig. 4c, bottom panels): Due to the quantum capacitance effect or the reduced DOS near its Dirac point in 2D graphene, we found that the $\Phi = -2.0$ V MS-DFT data could be reproduced with the $\mathcal{E}_0 = -0.20$ V/Å (rather than $\mathcal{E}_0 = -0.25$ V/Å) DFT calculation.

## 3. CONCLUSION

In summary, based on the MS-DFT framework, we formulated a first-principles theory to calculate the electric enthalpy of electrochemical interfaces or the non-equilibrium electrode-molecule adsorption energy. Specifically, we demonstrated that the internal energy of the non-equilibrium channel $C$ can be obtained by performing an MS-DFT calculation for the electrodes-only $L+R$ counterpart in addition to that for the full $L+C+R$ junction and identifying the difference between the two total energies. A subsequent DFT calculation for the channel $C$ under the electric field extracted from the $L+R$ electrodes-only MS-DFT calculation then provides the channel $C$ polarization, from which one can determine the electric enthalpy of the channel $C$ and by referencing to the equilibrium channel $C$ total energy the non-equilibrium electrode $L$-channel $C$ adsorption energy. Applying the developed theory to the Au(111)-single water and graphene-single water prototype electrochemical models, we calculated the electrode potential- and water configuration-dependent adsorption energies and found that Au and graphene electrodes show very different trends: On the Au electrode, the flat-water configuration was found to be the energetically most preferable conformation in the low electrode potential regimes and the H-down and H-up water configurations are stabilized only at large negative and positive electrode potentials, respectively. On the other hand, on the graphene electrode, the flat water-configuration was not stabilized and the direct transition from the H-down to H-up water configuration occurred as the potential sweeps from the negative to positive direction. Analyzing the flat-water configuration electronic structures in the two cases, we finally revealed that the origins of this difference is due to strong (weak) hybridization of the water O lone pair orbitals with the Au (graphene) states and their energetic pinning (shift) with the applied potentials. We believe that the non-equilibrium first-principles calculation theory presented here should prove to be a valuable tool in advancing the atomic-level understanding of electrochemical interfaces and accelerate the development of various energy and electronic devices.

## Methods

We implemented the developed theory within the SIESTA package[26] in combination with the in-house grid operation code[27,28]. We applied the MS-DFT formalism to an electrode-water interface model where the Au(111) or graphene working electrode $L$ is in contact with a single water molecule $C$ and is balanced by the reference electrode Au(111) $R$, as shown in Fig. 1b. To avoid the interactions between neighboring images of the water molecule, a supercell with dimensions of 9.993 Å × 8.655 Å for the Au electrode and 12.918 Å × 12.430 Å for the graphene electrode along the surface-parallel direction was used. The distance between the $L$ and $R$ electrode surfaces was set to 20 Å. According to previous studies[5,6], the ±0.4 V/Å electric field corresponds to an electrode potential of ±1.23 V with respect to the potential of zero charge under the assumption of an electrical double layer thickness of 3.0 Å, making our computational model a reasonable Helmholtz model compared with the experimental conditions. The centroid of the electric dipole moment of the water molecule $p_{H_2O}^V$ was set as the center of mass of the water molecule. All calculations were performed within the van der Waals density functional of optB88[29] to obtain accurate electrode-water interfacial properties. Double ζ-plus-polarization-level numerical atomic orbital basis sets were employed together with Troullier-Martins-type norm-conserving pseudopotentials[30]. A mesh cutoff of 600 Ry for the real-space integration was used, and 5 × 5 × 1 and 8 × 8 × 1 Monkhorst-Pack $k$-point grids of the Brillouin zone[31] were sampled for Au and graphene, respectively. The atomic geometries were optimized until the total residual forces were below 0.005 eV/Å using the conjugate gradient algorithm.

### ACKNOWLEDGEMENTS

This work was supported by the National Research Foundation of Korea (2023R1A2C2003816, 2022K1A3A1A91094293, RS-2023-00253716) and the Samsung Research Funding & Incubation Center of Samsung Electronics (No. SRFC-TA2003-01). Computational resources were provided by the KISTI Supercomputing Center (KSC-2022-CRE-0258).

### AUTHOR CONTRIBUTIONS

Y.-H.K. developed the formalism and oversaw the project. J. L. and R.G.L. implemented the method and J.L. and H.Y. carried out calculations. J.L, H.Y., and Y.-H.K. analyzed the computational results, and Y.-H.K. wrote the manuscript with input from J.L.

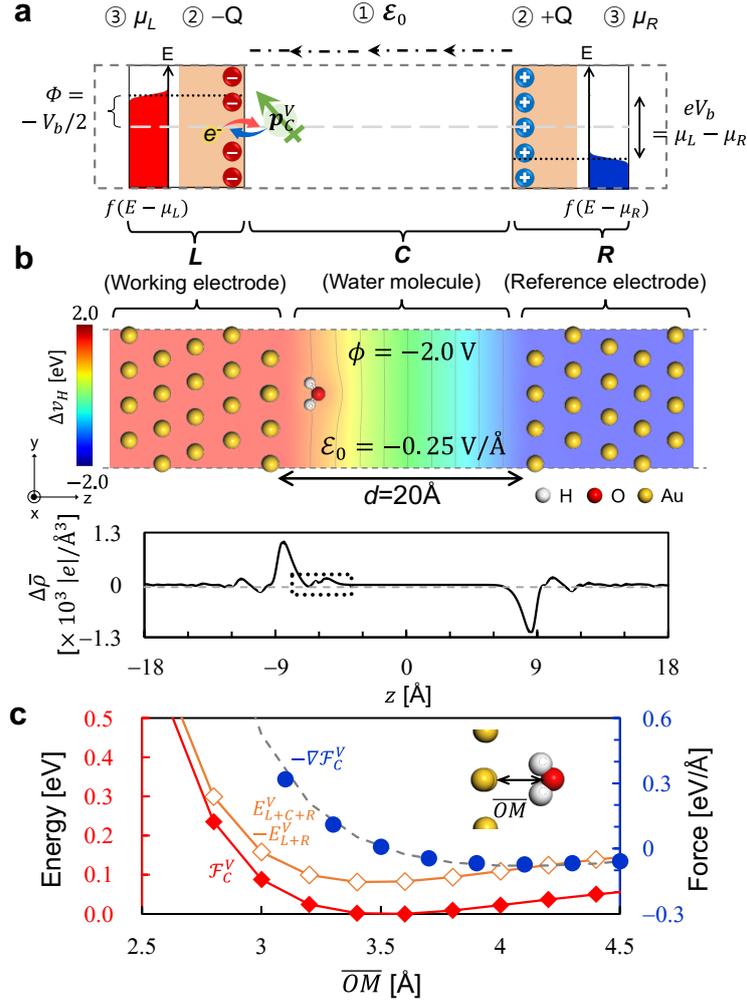

**Fig. 1 Formulation and validation of the electric enthalpy in electrochemical interface within the MS-DFT formalism. a** To describe the electrode potential $\Phi$-dependent interaction between a working electrode $L$ and a molecule $C$ with the dipole moment $p_C^V$ (green solid arrow), rather than introducing an electric field $\mathcal{E}_0$ across $C$ (①) or assigning charges on electrodes (②), MS-DFT explicitly introduces the bias voltage $V_b$ between $L$ and the reference electrode $R$, which are characterized by the Fermi functions $f$ with the electrochemical potentials $\mu_L$ and $\mu_R$, respectively (③). Electrode potential $\Phi$ is defined as $-V_b/2$. **b** The atomic structure of the Au(111)-single water-Au(111) interface model overlaid with the two-dimensional contour plot of the bias-induced electrostatic potential change $\Delta v_H$ (top) and the corresponding plane-averaged charge density difference or Landauer resistivity dipole $\Delta\bar{\rho}$ (bottom) obtained at $\Phi = -2.0$ V or $\mathcal{E} = -0.25$ V/Å. **c** The $\mathcal{F}_C^V$ (red filled diamonds) is shifted downward from $E_{L+C+R}^V - E_{L+R}^V$ (orange open diamonds) by the $\mathcal{E}_0 \cdot p_C^V$ contribution. The corresponding $-\nabla\mathcal{F}_C^V$ (dotted lines) curves match well with the z-component of the atomic forces $F_z$ (gray dashed line).

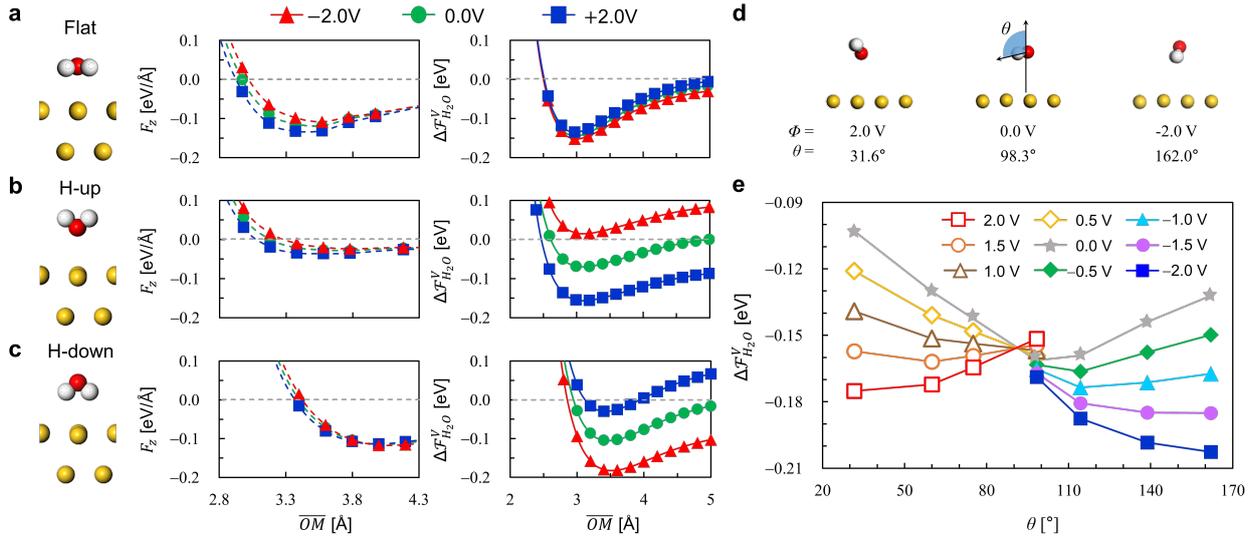

**Fig. 2 Non-equilibrium forces and adsorption energies of a single water molecule on metal surfaces.** For the **a** 'flat', **b** 'H-up', and **c** 'H-down' water configurations (left panels), the z-component atomic forces (middle panels) and adsorption energies $\Delta\mathcal{F}_{H_2O}^V$ (right panels) are shown as functions of the distance. Red triangles, green circles, and blue squares are the data obtained at $\Phi$ = -2.0 V, 0.0 V, and 2.0 V, respectively. **d** Relaxed configurations of a single water molecule on the Au(111) surface at $\Phi$ = -2.0 V (left), 0.0 V (middle), and 2.0 V (right). The angle $\vartheta$ is defined as the angle of the water plane from the Au surface-normal direction. **e** The $\Delta\mathcal{F}_{H_2O}^V$ variations as functions of $\vartheta$ at different electrode potentials.

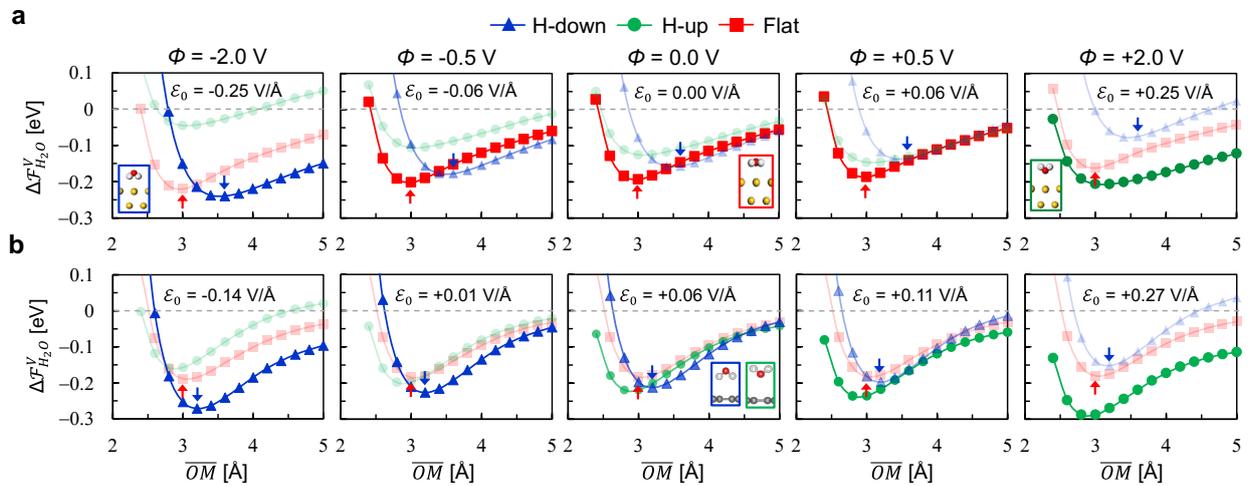

**Fig. 3 Comparison between electrode potential- and water configuration-dependent adsorption energies of a single water molecule on Au and graphene electrodes.** The bias-dependent adsorption energies of a single water molecule $\Delta\mathcal{F}_{H_2O}^V$ on the **a** Au and **b** graphene electrodes calculated for the H-down (blue triangles), H-up (green circles), and flat (red squares) configurations are presented as functions of the $\overline{OM}$ distance. In each panel, the electric field corresponding to the electrode potential is presented, and the curves for energetically less stable geometries are shown translucent. Additionally, the energetically most stable water structures are displayed in the insets of the panels for the lowest electrode potential cases.

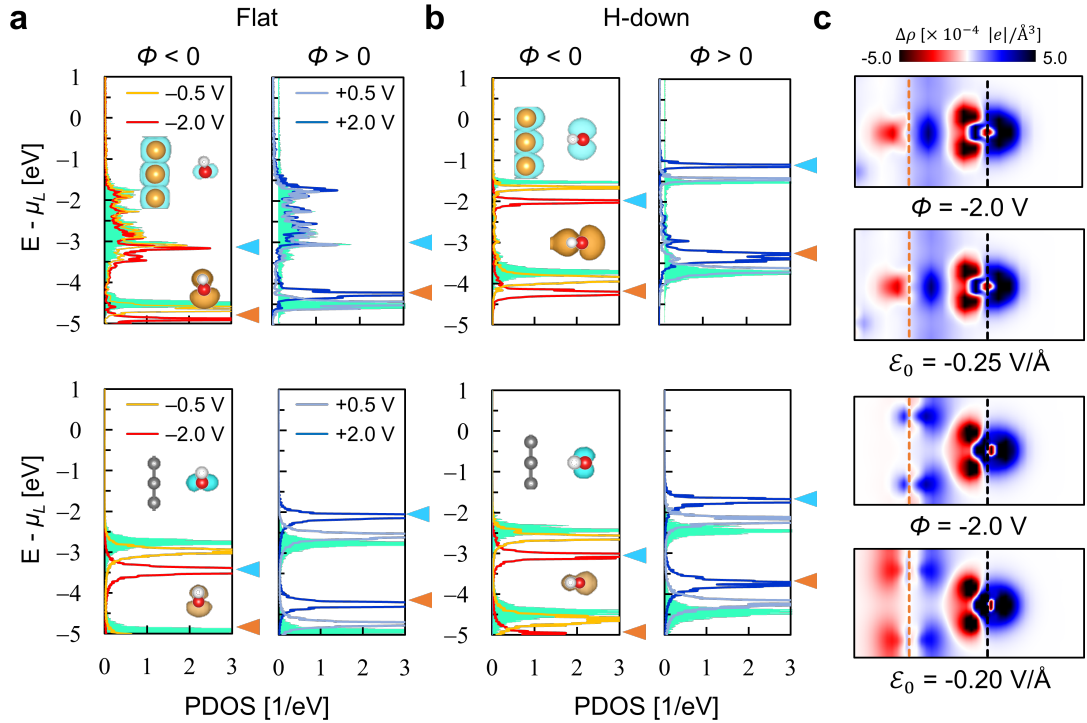

**Fig. 4 Electronic structures of the single water molecule on the electrified Au and graphene surfaces.** The PDOS of water in the **a** flat and **b** H-down configurations on Au (top panels) and graphene (down panels) calculated at $\Phi = -2.0$ V (red lines), -0.5 V (orange lines), 0.0 V (green shaded), +0.5 V (cyan lines), and +2.0 V (blue solid lines). Cyan and orange left triangles indicate the O lone pair and OH bonding states of the water molecule, respectively, and for $\Phi = -2.0$ V their local DOS extracted from the energy window of [- 0.1 eV, +0.1 eV] are overlaid on the atomic structures. The isosurface level is 0.01 states·Å$^{-3}$eV$^{-1}$. **c** The 2D contour plots of the $\Delta \rho$ obtained for the H-down water from constant-$\Phi$ MS-DFT calculations (top panels; -2.0 V for both Au and graphene) and constant-$\mathcal{E}$ DFT calculations (bottom panels; -0.25 V/Å for Au and -0.20 V/Å for graphene).

8